\documentclass[reprint, superscriptaddress, showkeys,amsmath,amssymb, aps]{revtex4-2}
\usepackage{graphicx}
\usepackage{dcolumn}
\usepackage{bm}
\usepackage{hyperref}
\usepackage{bm}
\usepackage{epstopdf}
\usepackage{epsfig}
\usepackage{bbold}
\usepackage{ulem}
\usepackage{xcolor}
\usepackage{notes2bib}

\begin{document}

\title{Quantum microscopy based on Hong-Ou-Mandel interference }
\author{Bienvenu~Ndagano}
\affiliation{School of Physics and Astronomy, University of Glasgow, Glasgow G12 8QQ, UK}
\author{Hugo~Defienne}
\affiliation{School of Physics and Astronomy, University of Glasgow, Glasgow G12 8QQ, UK}
\author{Dominic~Branford}
\affiliation{SUPA, Institute of Photonics and Quantum Sciences, Heriot-Watt University, Edinburgh, EH14 4AS, UK}
\author{Yash~D.~Shah}
\affiliation{School of Physics and Astronomy, University of Glasgow, Glasgow G12 8QQ, UK}
\author{Ashley~Lyons}
\affiliation{School of Physics and Astronomy, University of Glasgow, Glasgow G12 8QQ, UK}
\author{Niclas~Westerberg}
\affiliation{School of Physics and Astronomy, University of Glasgow, Glasgow G12 8QQ, UK}
\author{Erik~M.~Gauger}
\affiliation{SUPA, Institute of Photonics and Quantum Sciences, Heriot-Watt University, Edinburgh, EH14 4AS, UK}
\author{Daniele~Faccio}
\affiliation{School of Physics and Astronomy, University of Glasgow, Glasgow G12 8QQ, UK}
\date{\today}

\begin{abstract}
Hong-Ou-Mandel (HOM) interference, the bunching of indistinguishable photons at a beam splitter, is a staple of quantum optics and lies at the heart of many quantum sensing approaches and recent optical quantum computers. Here, we report a full-field, scan-free, quantum imaging technique that exploits HOM interference to reconstruct the surface depth profile of transparent samples. We demonstrate the ability to retrieve images with micrometre-scale depth features with a photon flux as small as 7 photon pairs per frame. Using a single photon avalanche diode camera we measure both the bunched and anti-bunched photon-pair distributions at the HOM interferometer output which are combined to provide a lower-noise image of the sample. This approach demonstrates the possibility of HOM microscopy as a tool for label-free imaging of transparent samples in the very low photon regime.
\end{abstract}

\maketitle

{\bf{Introduction}.}
The bunching of indistinguishable of photons at the outputs of a beam-splitter is the key signature of the Hong-Ou-Mandel (HOM) effect \cite{Hong1987}. Since the first demonstration, this effect has found many applications in various fields of quantum optics, from quantum state engineering \cite{Lee2012,Zhang2016}, quantum information processing \cite{Kok2007,Nagali2009} and quantum metrology \cite{Lyons2018,Scott2020}. In the context of quantum imaging, HOM interference has been exploited to engineer quantum states through post-selection with spatial light-modulators and single-pixel detectors, for multi-photon ghost-imaging \cite{Bornman2019,Bornman2019b}. One of the drawbacks of this approach is that it involves reconstructing the image one spatial mode at a time. To circumvent this limitation, there has been a growing interest in single-photon cameras  and related imaging opportunities \cite{Genovese2016,Moreau2019,GilaberteBasset2019} such as the characterisation of quantum correlations and entanglement \cite{Moreau2012, Edgar2012, Lubin2019,Ianzano2020,Ndagano2020}, ghost imaging \cite{Morris2015,Moreau2018}, quantum holography \cite{Devaux2019}, imaging with undetected photons \cite{Lemos2014}, imaging through noise \cite{Defienne2019,Gregory2020}, N00N-state imaging \cite{icfo,hugo21} and entanglement-enabled holography \cite{Defienne2021}. It is worth noting that while many of the schemes mentioned have relied by CCD technology, next-generation single-photon avalanche photodiode cameras with high temporal resolution, high pixel count and high frame-rates, are poised to enable even more exciting pathways in quantum imaging \cite{Morimoto2020,Morimoto2020a}.\\
In the context of HOM interferometry, Chrapkiewicz \textit{et al.} have demonstrated the measurement of the spatial structure of a single photon using an intensified CMOS camera \cite{RadoslawChrapkiewicz2016}; spatial indistinguishability between the photons is altered by a digital hologram, leading to varying degree of photon-bunching, revealing the shape of the hologram. It is however, worth noting that this demonstration evaluated the indistinguishability of two single-mode twin beams that were filtered with single-mode fibres. A similar approach was taken by Ibarra-Borja \textit{et al.} for full-field quantum optical coherence tomography \cite{Ibarra-Borja2020}; by scanning the delay between twin photons, they reconstruct the image of a sample using an intensified CCD camera. Although multi-mode HOM sensing has been shown by Devaux \textit{et al.} \cite{Devaux2020} where they replaced single-pixel detectors with two electron-multiplying CCD cameras for coincidence counting, we note no report of a multi-mode  HOM imaging technique. Such a scheme will have the potential benefit of higher spatial resolution over single-mode illumination. In addition, a wide-field technique would enable bio-imaging without any scanning parts, significantly reducing acquisition times whilst still operating in the very low photon regime.\\
{{Rather than competing with classical imaging techniques, quantum imaging offers complementary benefits and provides additional opportunities such as low light imaging or more robust interferometric sensing approaches. The former may be crucial for delicate photoactive samples, whereas the latter derives from measuring interference via photon correlations, making it robust to perturbations that extinguish classical interferometric fringes.}}\\
%
\begin{figure}[t]
	\centering\includegraphics[width= 7cm]{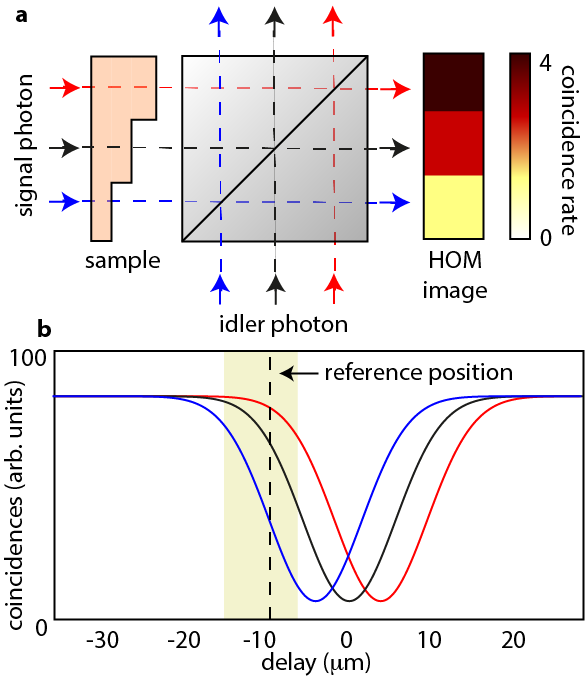}
	\caption{\label{Figure1} \textbf{Principles of Hong-Ou-Mandel imaging.} {\bf{a}}. In a Hong-Ou-Mandel interferometer, the paths of two indistinguishable photons, signal and idler, overlap on a 50/50 beam-splitter. The signal photon traverses a transparent sample of varying thickness, and the outputs of the beam-splitter are, colour-wise, measured in coincidences.  {\bf{b}}  When adjusting the idler delay to the reference position (dashed line), the three paths map different coincidence probabilities, allowing one to obtain a contrast image of the transparent sample for a range of depths indicated by the shaded area. This can be used to reconstruct the depth thickness variation across the sample. }
\end{figure}
\begin{figure}[h]
	\centering\includegraphics[width= 8cm]{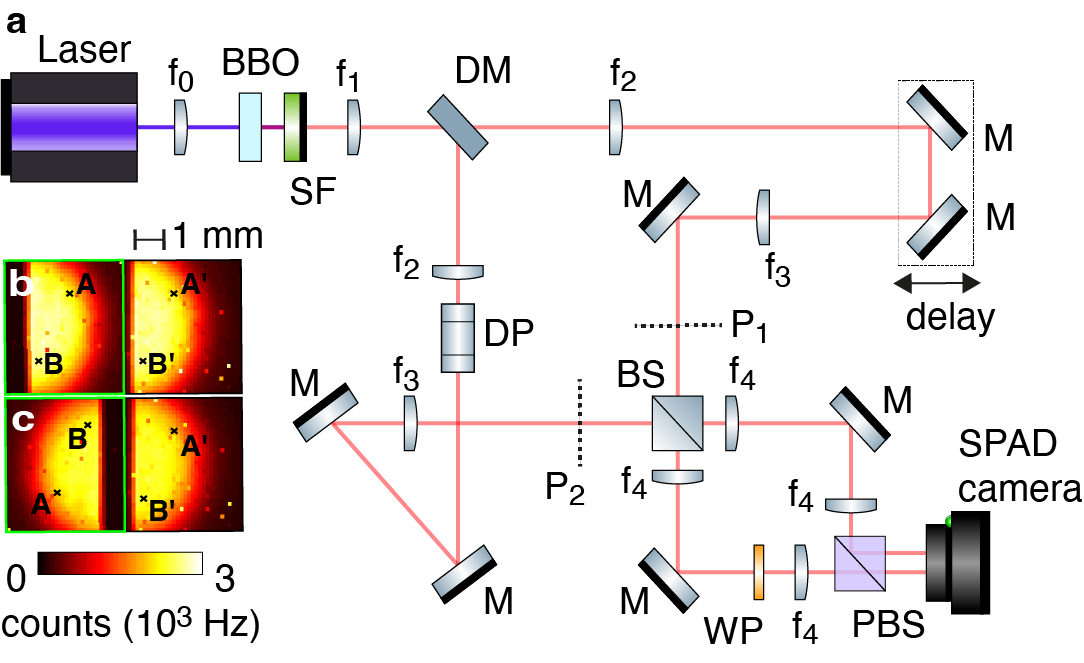}
	\caption{\label{Figure2} \textbf{Hong-Ou-Mandel imaging setup.} {\bf{a}}. A 0.7-mm diameter (1/$e^2$) collimated pump beam from a 347-nm pulsed laser with a 100-MHz repetition rate, is focused into a 0.5-mm-long $\beta$-barium borate (BBO) crystal where photons pairs are generated through type-I spontaneous parametric down-conversion. Signal and idler photons and separated in the far-field of the BBO crystal using a D-shaped mirror (DM). The signal photon is sent through a delay line where the optical path can be adjusted using a motorised stage with a 1 micron step size. The idler photon propagates through a dove prism (DP) that performs an inversion around the $x$-axis. The plane of the D-shaped mirror is relayed to planes P1 and P2 using two identical $4f$ imaging systems with a $2\times$ magnification. Subsequently, planes P1 and P2 are overlapped and imaged onto the SPAD camera, using two identical $4f$ telescopes with a $1\times$ magnification. {\bf{b}}. Intensity image acquired by the SPAD camera, where spatial positions A(B) and A'(B') map to photon paths from the two output ports of the BS. \textbf{c} We apply a $\pi$ rotation to one half of the image (enclosed in the green box) such that signal and idler photons measured A(B) and A'(B') respectively, are spatially anti-correlated. f$_0$ = 300 mm, f$_1$ = 100 mm, f$_2$ = 100 mm, f$_3$ = 200 mm, f$_4$ = 150 mm, SF = spectral filter, BS~=~50:50 beam-splitter, PBS = polarisation beam-splitter, WP = half-wave plate, M = mirror. }
\end{figure}
%
\begin{figure}[t]
	\centering\includegraphics[width= 6.8cm]{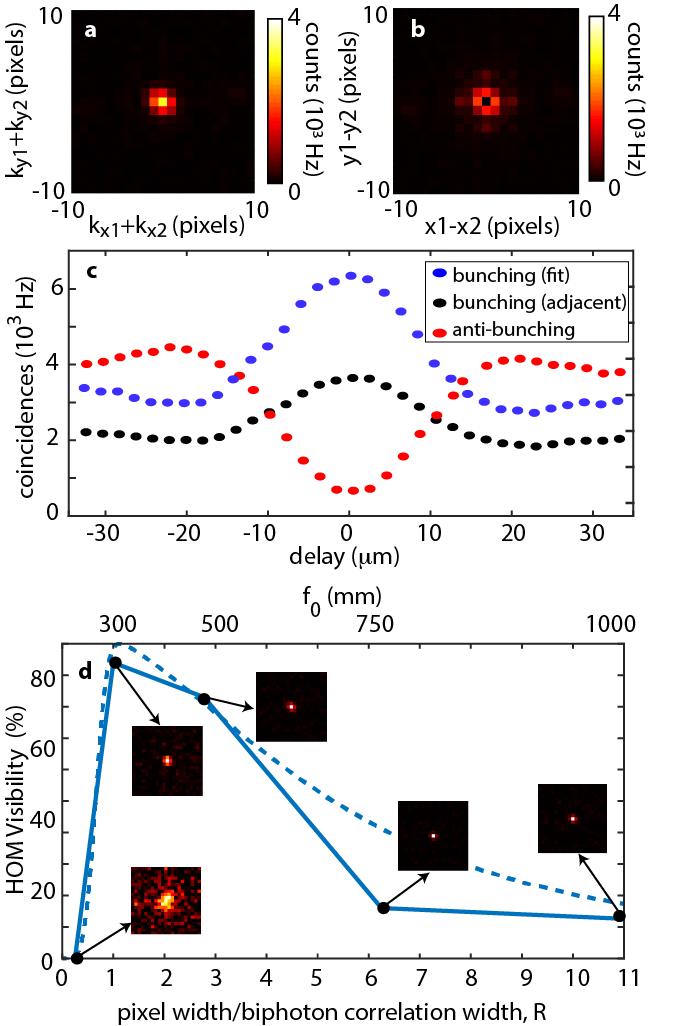}
	\caption{\label{Figure3} \textbf{Hong-Ou-Mandel sensing with a SPAD camera.} Following the reconstruction of the two-photon spatial distribution, its projection onto the \textbf{a}, sum-coordinates reveals a correlation peak corresponding to the numbers of photons measured at anti-correlated positions (anti-bunching). \textbf{b} Similarly, the projection onto the minus-coordinates evaluates the number of photons bunching. The fraction of photons bunching can be evaluated by either fitting a peak, or measuring the signal from adjacent pixels. \textbf{c} By scanning the delay stage, we measure the Hong-Ou-Mandel interference. The legend indicates the anti-bunching results (HOM interference dip) and the bunching results (HOM interference peak) under the two approximations tested (error bars indicating standard error of the mean, are the same size or smaller than dots). At each delay position, we acquire acquire and analyse on average 19 million intensity frames.  \textbf{d} HOM interference visibility versus {focal length of the PDC-pump focusing lens (top axis) and ratio, R, of the camera pixel width to the biphoton correlation width (bottom axis). The highest visibility is observed when the pump is focused such that the two-photon correlation width is equal to the camera pixel pitch. The insets show the normalised sum-coordinate projections of the JPD for different pump lens focal lengths (plotted over 10x10 pixels). The dashed curve is a parameter-free model used to calculate the HOM dip visibility for our experiment}.}
\end{figure}
%
Here, we demonstrate a full-field, scan-free, quantum imaging technique enabled by HOM interference. The scheme exploits the fact that group velocity delays along the rising (or falling) edge of the HOM interference signal, have a 1:1 mapping to coincidence rate. We spatially resolve the HOM interference across multiple spatial modes by reconstructing the two-photon spatial joint probability distribution at every pixel position of a single photon avalanche diode (SPAD) camera, from which we extract both the photon-bunching and photon-anti-bunching information. We can choose to use the latter to obtain the depth profile of samples such as a pattern of clear acrylic sprayed over a microscope slide with an average depth of $\sim 13$  $\mu$m or a pattern etched on a glass substrate of  $\sim 8$  $\mu$m  depth. In both cases, we observe that the depth profile of the sample is not accessible through direct intensity measurement. However, when observed with the HOM imaging system, we obtain a contrast image that does reveal the structure of the object. Image resolution is enhanced via a standard 2x2 camera raster scanning technique whilst noise in the image is reduced by combining information from both the bunched and anti-bunched photons and based on ideas recently introduced in Ref.~\cite{Scott2020} . The combination of these approaches and the high frame rates of the SPAD camera, allow efficient imaging of micron-sized features at very low-photon levels.\\
%
\begin{figure}[ht]
	\centering\includegraphics[width=8cm]{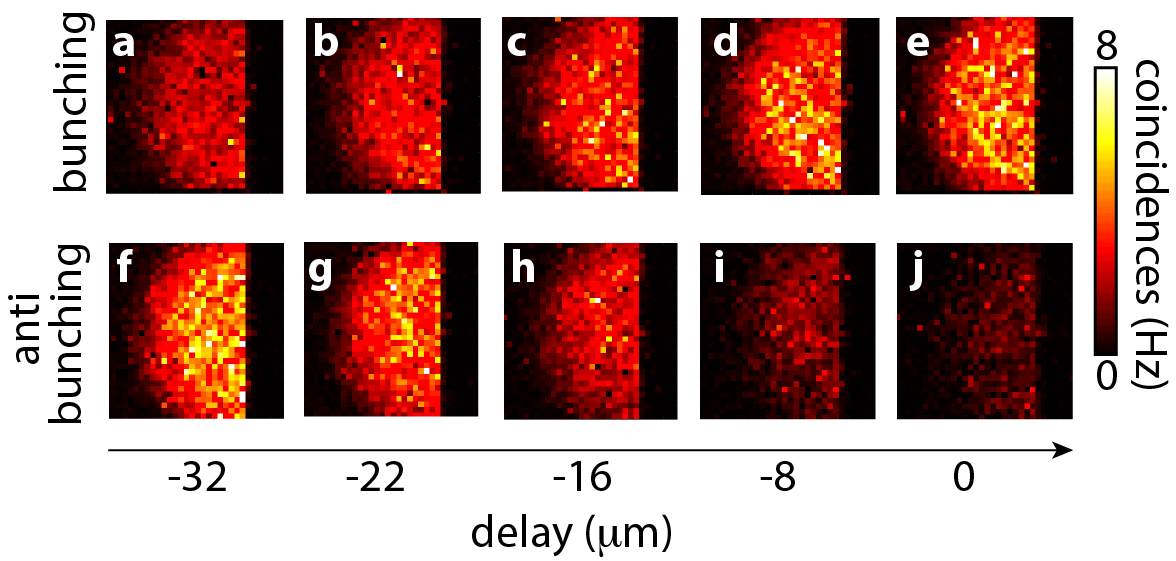}
	\caption{\label{Figure4} \textbf{Full-field Hong-Ou-Mandel sensing.} 32x32 pixel spatial images of photon coincidences arising from bunching (\textbf{a}-\textbf{e}) and anti-bunching (\textbf{f}-\textbf{j}), reconstructed as a function of spatial delay between signal and idler photons. The bunching images are obtained by measuring coincidences across adjacent pixels. At each delay position, the two-photon distribution is reconstructed from an average of 19 million intensity frames. Pixel pitch is 150 $\mu$m.}
\end{figure}
%
%
{\bf{Concept}.}
 The idea behind our Hong-Ou-Mandel imaging technique is shown in Fig.~\ref{Figure1}. The paths of two indistinguishable photons are overlapped onto a 50/50 beam-splitter. A (signal) photon travels through a sample with a varying thickness, while the other (idler) photon does not. For each of the three colour-coded trajectories, the signal photon incurs different group velocity delays, leading to different arrival times with respect to the idler photon on the matching trajectory. Coincidence measurements on each of the colour-coded paths at the outputs of the beamsplitter, show that the two-photon interference signals are shifted with respect to each other. By recording the coincidence rates in each of the colour-coded paths, one can obtain a contrast image of the sample. Figure~\ref{Figure1}b illustrates the mapping between spatial delay and coincidence probability when measuring at a fixed delay reference position that can be exploited to reconstruct the relative depth profile of the sample from its HOM image without any need for scanning the HOM interferometer delay. \\
 {{Compared to classical interferometric or phase imaging approaches, HOM interference does not require phase-stability of the setup yet can still achieve similar 1-10 nm depth resolution sensitivity \cite{Lyons2018}. The simple approach described here that operates at a fixed delay position at the HOM dip edge (rather than scanning the interferometer across the dip) can achieve and axial (depth) resolution of 100 nm   even on a mobile platform \cite{rotating_HOM}. Differently from classical interferometry, the ``axial field of view'' (distance over which the sample depth can be resolved) can be significantly larger than the optical wavelength and is determined by the half-width of the HOM dip (which in turn is fixed by the SPDC spectral bandwidth). This is of order $\sim20$ $\mu$m in our experiments.}}\\
\noindent\textbf{Experimental setup}. The layout of the HOM imaging system is depicted in Fig.~\ref{Figure2}a (See Methods for full details). Signal and idler photon pairs are generated via spontaneous parametric down-conversion and are spatially separated in the far-field using a D-shaped mirror.  Both signal and idler propagate through identical 4$f$-imaging systems that relay the far-field to planes P$_1$ and P$_2$ -- the sample to be imaged is placed in P$_2$. The image of planes P$_1$ and P$_2$ are overlapped using a 50:50 beam-splitter (BS) and imaged, using identical imaging systems onto a SPAD camera (SPC3 from MPD) with an array of 32 $\times$ 64 pixels, an 80\% fill-factor and a 150-$\mu$m pixel pitch that can acquire up to 96 kframes/second.\\
The two outputs from the BS are shown in (Fig.~\ref{Figure2}b, where pixel positions A(B) and A'(B') map to photon paths in the two outputs of the BS. In the event of bunching, pairs of photons would be detected at either A(B) or A'(B'). Meanwhile in the event of anti-bunching, one photon in a pair would be detected at A(B) and the other at A'(B). By applying a $\pi$-rotation on one of the output arms, we note that spatially correlated pair-detection is indicative of bunching, while spatially anti-correlated pair-detection indicates anti-bunching.\\
%
\begin{figure*}[th]
	\centering\includegraphics[width=0.9\linewidth]{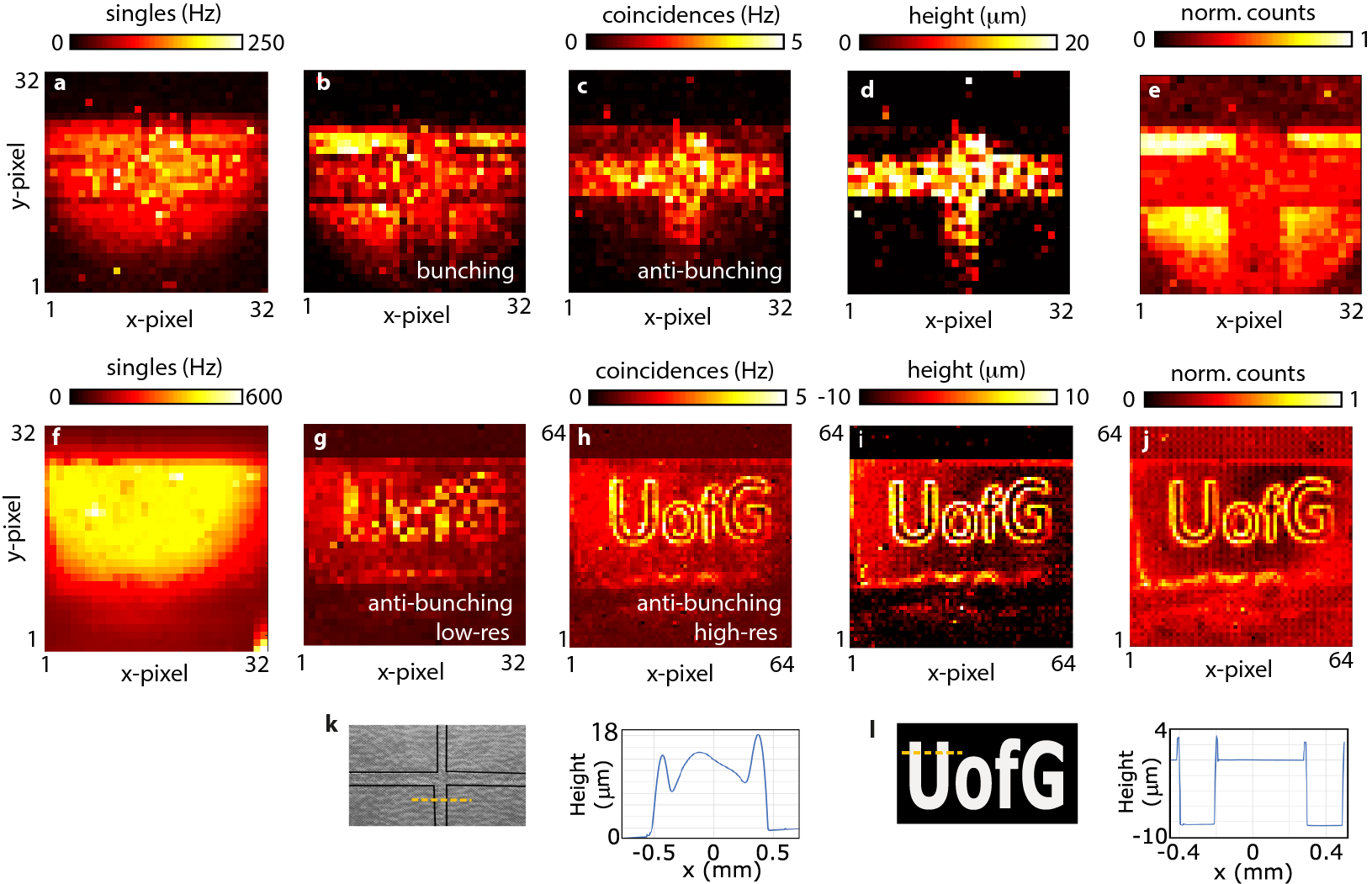}
	\caption{\label{Figure5} \textbf{Full-field Hong-Ou-Mandel imaging. a} The intensity image as captured by the SPAD camera of a clear acrylic cross-pattern on a microscope slide.  \textbf{b} Shows the bunching (from adjacent pixels) and \textbf{c} anti-bunching coincidence maps. \textbf{k} shows an image of the sample and the profile measurement taken along the yellow dashed line - average height is 12.9 $\mu$m. \textbf{d} The number of coincidence events is converted to a height measurement relative to the surface of the microscope slide. \textbf{e} The combination of weighted bunching and anti-bunching images produces a normalised image with reduced noise. \textbf{f} Intensity image of the letters `UofG' etched into a glass substrate, to a depth of 8.36 $\mu$m.   \textbf{g} Shows the anti-bunching (from adjacent pixels) at the native 32x32 camera resolution and \textbf{h} is 64x64 pixel super-resolved image obtained by 2x2-raster scanning the camera.  \textbf{l} shows an image of the sample and the profile measurement taken along the yellow dashed line. \textbf{i} Depth map of the sample obtained from the coincidence map in \textbf{h}. \textbf{j} The combination of weighted bunching and anti-bunching images produces a normalised image with reduced noise. Pixel pitch is 150 $\mu$m.}
\end{figure*}
%
\noindent\textbf{Hong-Ou-Mandel sensing}. At the plane of the SPAD camera, we reconstructed the signal ($\mathbf{s}$) and idler ($\mathbf{i}$) joint probability distribution (JPD), $\Gamma(\mathbf{r_s},\mathbf{r_i})$, using the following model \cite{Defienne2018}:
\begin{equation}	
	\label{eq: JPD intensity}
	\Gamma(\mathbf{r_s},\mathbf{r_i}) = \frac{1}{N} \sum_{l=1}^N I_l(\mathbf{r_s})I_l(\mathbf{r_i}) - \frac{1}{N^2} \sum_{m,n=1}^{N} I_m(\mathbf{r_s})I_{n}(\mathbf{r_i}),
\end{equation}
where $I_l(\mathbf{r}) \in \{0,1\}$ is the binary value returned by the SPAD sensor for a pixel at location $\mathbf{r}$ in the $l^{th}$ frame. Over the acquisition time of the camera, set here to 10~$\mu$s, the sensor may measure photons from multiple pairs. The intra-frame correlation in Eq.~(\ref{eq: JPD intensity})  (first term on the right-hand side) estimates the number of coincidences from photons belonging to the same pair (genuine coincidences), as well as those from different pairs (accidental coincidences). The latter are estimated and removed from the reconstructed JPD by subtracting the inter-frame correlations (second term on the right-hand side). \\
As highlighted above in the experimental configuration, anti-bunching events are registered as pair-detections from spatially anti-correlated photons and are estimated by projecting the JPD onto the sum-coordinates, Fig.~\ref{Figure3}a. The height of the measured correlation peak indicates the number of reconstructed anti-bunching events. The number of bunching events can be extracted from pair-detections of spatially correlated photons, i.e. we project the JPD onto the minus-coordinates as show in Fig.~\ref{Figure3}b. In this case, we do not have a correlation peak because the pixels on the SPAD camera are not able to photon-number-resolve. Therefore, it is not possible to directly measure events of two photons incident on the same pixel.\\
We then characterised the HOM dip (peak) from two-photon interference by scanning the signal arm delay stage and evaluated the number of anti-bunching and bunching events at each pixel, as shown in Fig.~\ref{Figure3}c and obtain a HOM dip with a visibility of $88\pm2\%$ (red circles). To then estimate the number of bunching events, we can either fit a Gaussian peak to estimate the correlation amplitude of the central pixel (blue circles, $81\pm7\%$ visibility) or simply average over the four nearest neighbours to  the central pixel (black circles, $60\pm7\%$). The latter has lower visibility as one may expect but was the preferred option in the following results as it relies only on measured data (see Methods).\\
\noindent\textbf{Tailoring the two-photon correlation}. 
The spatial width of the two-photon correlation provides a measurement of the average mode width and plays a key role in optimising the HOM visibility measured by the camera.  The correlation width can be controlled by the pump diameter as this will determine the number of modes and the divergence (k-vector spectrum) of the SPDC. Figure~\ref{Figure3}d shows the measured HOM dip visibility as we vary the pump beam diameter on the SPDC crystal by changing the pump beam focusing lens, $f_0$ {(upper horizontal axis) or equivalently, as a function of the ratio, R, of pixel width to the biphoton correlation width (lower horizontal axis, see Methods for more details). }
 If the correlation width is much smaller than the camera pixel pitch (large focal lengths, loose pump focusing), then many modes will overlap at the same pixel. If the correlation width is very broad (short focal length, tight pump focusing) then each single mode spreads across many pixels and will overlap with other modes. In both cases, HOM visibility is lost.  Instead, the highest HOM interference visibility is obtained when each pixel acts a single mode detector, i.e. when the correlation width  is of the same order of the camera pixel size of 150 $\mu$m (in our case, this corresponds to $f_0=300$ mm). {{The insets in Fig.~\ref{Figure3}d show the measured biphoton distributions. The dashed curve shows a model for the HOM dip visibility for our experiment with no free parameters and confirms the role and importance of tailoring the spatial mode content of the quantum light field on the camera.}}\\
\noindent\textbf{Full-field Hong-Ou-Mandel interferometry}. 
%
%
Figure~\ref{Figure4} shows the spatially resolved bunching and anti-bunching coincidence maps for 5 different delays over one half of the sensor, the other half simply being a symmetric image and which contains the same information. As the interferometer spatial delay tends to zero, the number of bunching events increases to a maximum, while the anti-bunching events tend to zero. Crucial to our HOM imaging technique, this full-field HOM measurement shows the direct pixel-wise-resolved mapping between spatial delay and number of coincidence events. \\
{\bf{Hong-Ou-Mandel imaging}.}  A first sample was prepared by spraying a layer of clear acrylic on a glass substrate forming a cross-pattern with a depth of $12.9$ $\mu$m, averaged along the dashed line (inset above Fig.~\ref{Figure5}b), as measured with a profilometer. Figure~\ref{Figure5}a shows the intensity (photon counts)  image as recorded by the SPAD camera. The sample itself, as expected, is not visible although the edges are barely visible likely due to scattering that leads to an effective loss. The HOM images are extracted from the JPDs reconstructed from a total of 130 million intensity frames (total acquisition time of 37 minutes) and are shown in Figs.~\ref{Figure5}b and \ref{Figure5}c, for the bunching and anti-bunching coincidence events respectively.  These images show a clear contrast between the acrylic surface and its surroundings with
the anti-bunching and bunching images that appear as contrast reverse images of each other, due to conservation of probability (i.e. with small discrepancies due to noise and any losses).\\
We use the spatially resolved coincidence counts resolved versus delay in 4 $\mu$m steps (of which 5 delays are shown in Fig.~\ref{Figure4}) as a depth estimator and obtain from Fig.~\ref{Figure5}c, the  relative depth profile of the sample shown in Fig.~\ref{Figure5}d. 
From this, we estimate the average thickness of the layer of acrylic to be $14.5\pm6$ $\mu$m, in relatively good agreement with the ground truth measured value. We note the large error (standard deviation) around the average value that is due to the fact that the sample itself is not uniform but also due to the obvious noise in images. However, recent work showed that photon number resolving information can increase precision in (lossy) conventional HOM sensing  by combining bunching and anti-bunching signals \cite{Scott2020} (see Methods). A significantly improved image is then retrieved, Fig.~\ref{Figure5}e, which has a variance that is 3.4x smaller compared to the direct measurement Fig.~\ref{Figure5}c in the cross region of the image. \\
A second sample was fabricated by etching the letters `UofG' onto a glass substrate to a depth of 8.36~$\mu$m  (measured with a profilometer, inset to Fig.~\ref{Figure5}g showing also profile along the yellow dashed line) and width of the lettering etch, 230 $\mu$m. Figure~\ref{Figure5}f shows the intensity image that does not reveal any details about the shape of the etched sample. Figure~\ref{Figure5}g shows the anti-bunching coincidence image at the camera native 32x32 pixel resolution. The `UofG' pattern is visible but is strongly under-resolved. We therefore also performed a 2x2 raster scan of the camera so as to increase the pixel resolution by a factor 4x (by simple `shift and add' of the four raster-scan images \cite{raster}): the sample is now clearly visible in the anti-bunching HOM image in Fig.~\ref{Figure5}h. {{We note that the cross/UofG patterns have positive/negative thickness step changes with respect to the substrate - our technique does not require prior knowledge of this but rather, the changes in coincidence counts will be positive/negative respectively, directly indicating whether features on the substrate are positive or negative thickness variations.}
We also observe high counts at the edges of the etched regions. This is due to the very sharp profile of the etching (compared to the relatively smooth edges of the acrylic sample) that leads to edge-diffraction along the contours of the letters. This diffraction introduces additional transverse wave-vector components on the transmitted photons, which reduces the indistinguishability between signal and idler photons and a higher coincidence rate with respect to the surrounding non-etched parts. Figure~\ref{Figure5}i shows the retrieved depth image from the high-resolution coincidence image from which we estimate an etching depth of $8.2 \pm 1.2$ $\mu$m, in very good agreement with the ground truth reference measurement and also shows a relatively low error that is in line with typical HOM measurements that use single point detection. As for the case of the cross, a weighted combination of the bunching and anti-bunching data has a reduced noise with a reduced variance of 1.6x inside the letters (Fig.~\ref{Figure5}j). \\
%
{\bf{Conclusions}.} HOM interference can be used in full-field imaging to directly retrieve spatially-resolved depth profiles of transparent samples. Access to both bunching and anti-bunching images can be used to also assess losses and in turn reduce the noise variance in the images by up to nearly an order of magnitude in the best case. \\
To put these measurements into context, we note that the average coincidence values are of order 1 Hz for each pixel, which at 60 kframes/second implies roughly only 1 frame every 60 detecting an actual photon pair. If we consider the camera photon detection probability (80\% fill factor and 6\% quantum efficiency), this corresponds to an actual average photon pair flux in the interferometer of $\sim400$ photon pairs/second at each pixel or $\sim7$ photon pairs/frame. This is extremely low yet this still allows us to retrieve clear images with $\mu$m-level absolute precision in the depth measurements. Photon density and total illumination on the sample are also often quoted as a concern for some bio-imaging applications. The results shown here were performed in a regime in which $\sim10^5$ photons at each pixel are required to illuminate the sample for the total minimum exposure time of $\sim20$ minutes (if running the camera at 96 kframes/second). This can be possibly be further reduced by more than order of magnitude by improving the camera technology as indeed, photon pair detection scales quadratically with the camera quantum efficiency. HOM imaging therefore provides an opportunity for example for label-free bio-imaging {{or imaging of photoinduced effects that may require very low photon fluxes. Then, the illumination source itself does not modify the biological sample or at least provides controlled modification at the level of single photons  \cite{bio_quantum}}}.\\
 {{New generation asynchronous read-out SPAD cameras that can operate at $\sim50$ Mhz rates \cite{instruments4020014,fast_SPAD} and have already been used to increase up to nearly video frame rate other challenging imaging feats such as non-line-of-sight imaging \cite{fast_nlos} would also provide a 1000-fold decrease in acquisition times, potentially leading to video frame rate imaging capability.}} {{An increase in count-rates would also allow to apply high-precision HOM sensing approaches  with 10-100 nm sensitivity and competing with classical interferometric approaches albeit with the advantages of better stability and better axial filed-of-view that extends over tens of microns rather than a few hundred nm. }}\\
The methods demonstrated here can also be transferred e.g. to quantum optical coherence tomography that essentially also leverages photon bunching to detect multiple interfaces and layered structures. \\

\section*{\uppercase{Data availability}} 
The experimental data and codes that support the findings presented here are available from http://dx.doi.org/10.5525/gla.researchdata.1241.


\section*{\uppercase{Acknowledgements}} 
The authors thank M. Cromb for fruitful discussions on the interpretation of the results. The authors acknowledge financial support from the UK Engineering and Physical Sciences Research Council (grants EP/R030413/1, EP/M01326X/1 and EP/R030081/1) and from the European Union's Horizon 2020 research and innovation programme under grant agreement no. 801060. D.F. acknowledges support from the Royal Academy of Engineering Chair in Emerging Technologies programme. H.D. acknowledges support from the European Union's Horizon 2020 research and innovation programme under the Marie Sklodowska-Curie grant agreement No. 840958. N.W. acknowledges support from the Royal Commission for the Exhibition of 1851.

\section*{\uppercase{Authors contributions}} 
D.F. conceived the concept and supervised work. B.N., H.D., A.L. and D.F. conceived and discussed the experimental setup. B.N. performed the experiment. Y.D.S. micro-fabricated the etched sample. D.B., N.W. and E.M.G. applied noise reduction approaches. All authors contributed to the analysis of the results and the manuscript. 

\section*{\uppercase{Corresponding author}} 
{Daniele.Faccio@glasgow.ac.uk}

\section*{\uppercase{Competing interests}} 
The authors declare no competing interests.

\section*{\uppercase{Methods}} 
{\bf{Experimental layout.}} Signal and idler photon pairs are generated via type-I spontaneous parametric down-conversion (SPDC) in a 0.5-mm long $\beta$-barium borate (BBO) crystal. The generated photons are spatially separated in the far-field of the BBO crystal using a D-shaped mirror. The signal photon travels through a delay line where the path length can be adjusted using a motorised translation stage. The idler photon travels through a fixed length path where a dove prism performs an image inversion in the transverse plane. Both signal and idler propagate through identical 4$f$-imaging systems that relay the far-field of the BBO crystal (image plane of the D-shaped mirror) to planes P$_1$ and P$_2$ -- the sample to be imaged will subsequently be placed in the latter plane. The image of planes P$_1$ and P$_2$ are overlapped using a 50:50 beam-splitter (BS) and imaged, using identical imaging systems in both output ports, onto a SPAD camera (SPC3 from MPD) with an array of 32 $\times$ 64 pixels, a $\sim$78\% fill-factor and a 150-$\mu$m pixel pitch. The camera has a quantum efficiency of $\sim$9\% at the photon pair wavelength (694 nm), a nominal frame-rate of 96 kframes/second and a dark-count rate of 0.14 counts/pixel/second. For most of the experiments, the camera was operated at 60 kframes/second due to software/computer limitations.\\
{{{\bf{Choice of pump laser: }}The camera is internally triggered implying that a CW laser can also be used for SPDC generation, ideally with as short a wavelength as possible so as to benefit from the higher quantum efficiency of the camera at shorter SPDC wavelengths (e.g. $\sim10\%$ at 700 nm compared to  $\sim4\%$ at 800 nm. However, the choice of a pulsed laser source also provides photons with a broader bandwidth, i.e. a narrower HOM interference dip and consequently larger variations in the coincidence counts for a given sample thickness.}}\\
{{{\bf{Lateral spatial resolution: }} The lateral spatial resolution is set by the biphoton correlation width.  As shown in Fig. 3. this needs to be matched to the pixel size for optimum HOM visibility - then the field-of-view and lateral spatial resolution are set by the camera in a ?lens-less configuration?: in our measurements, this gives a FoV (determined by the camera chip size) of 4.8x4.8 mm$^2$ with 0.15 mm resolution (that is improved to 0.75 mm by raster scanning). One could alternatively focus the correlation width to a diffraction limited spot onto a sample to increase resolution, and then magnify the SPDC to mode match the pixel size onto the SPAD. The FoV would then be 4.8x4.8/M2 mm$^2$  where M is the magnification factor of the system and resolution now becomes the width of the diffraction limited spot.}\\
{\bf{Joint Probability Distribution measurements.}} 
At each delay position, we reconstructed the JPD from a total of 19 million intensity frames and measured a HOM dip with a visibility of $88\pm2\%$. To then estimate the number of bunching events, we used two approaches. The first consists in fitting a Gaussian peak to the minus-coordinates projection and use the peak as the estimate. Using this approach, we measured a HOM peak visibility of $81\pm7\%$, comparable to that obtained for HOM dip. The second approach exploits the fact that (i) the two-photon correlation width is larger than one pixel and (ii) the SPAD camera has a relatively high fill-factor (80\%). Thus, we can estimate the number of bunching events where photons are incident on adjacent pixels, i.e. { coincidences are generated from the conditional distribution $\Gamma(\mathbf{r}|\mathbf{r}+\Delta\mathbf{r})$, where $\Delta\mathbf{r}$ is a transverse shift by a single pixel and averaged over the four nearest neighbour pixels}. The visibility measured in this case is $60\pm7\%$. This lower value is to be expected given that adjacent pixels capture different spatial modes, thus increasing photon distinguishability.  \\
{{\bf{HOM dip visibility dependence on biphoton correlation width.}} 
The biphoton correlation width scales linearly with the pump laser beam area, which in turn scales quadratically with $f_0$. This is used in Fig.~\ref{Figure3}d to estimate the biphoton correlation width and thus map $f_0$ (upper horizontal axis) to R (lower horizontal axis) for all $f_0$ starting from the measurement at  $f_0=300$ mm that gives a biphoton correlation width equal to the camera pixel width (R=1) - all other R values are consequently scaled quadratically with $f_0$.\\
{\bf{Predicting the HOM dip visibility as a function of R:}}\\
The biphoton correlation function can be expanded in terms of Hermite-Gauss modes. We limit our 1D toy model to an equal superposition of the fundamental and first order modes, as this captures both cases of spatially symmetric and antisymmetric states which lead to bunching and antibunching, respectively \cite{padua2003}). A calculation similar to Ref.~\cite{padua2003} then yields the coincidence probability between positions $x_1$  and $x_2$ on the two detector-halves, respectively, at a delay $\delta$ of $P_C(x_1,x_2,\delta) = \psi^2_{+}(x_1,x_2)+\psi^2_{-}(x_1,x_2)+2\psi_{+}(x_1,x_2)\psi_{-}(x_1,x_2)\exp\left[-\delta^2/\left(2\Sigma^2\right)\right]$, where $\Sigma$ is the HOM dip width and $\psi_{\pm}(x_1,x_2) = \pm\mathcal{N}e^{-(x_1+x_2)^2/2\sigma^2_\text{corr}}e^{-\beta^2(x_1-x_2)^2/w^2}\left[1\pm \frac{\sqrt{2}}{\sigma_\text{corr}}(x_1+x_2)\right]$. Here $\sigma_\text{corr}=\sqrt{L_c/\beta^2 k_p}$ is the correlation width, $k_p$ and $w$ is the pump wavenumber and width, respectively, $L_c$ is the crystal length, $\mathcal{N}$ is a normalisation constant, and $\beta$ is a scaling constant to account for diffraction during the propagation between crystal and detector. Keeping the illumination area ($w/\beta$) constant for different pump beam sizes ($w$) leads to an effective change in the correlation length ($\sigma_\text{corr}=\sqrt{L_c/\beta^2 k_p}$) -- for instance $\beta^{-1} \simeq \{50, 25, 15\}$ for focal lengths $f_0 = \{150, 300, 500\}\text{ mm}$. The probability of detection at the $i^\text{th}$ and the $j^\text{th}$ pixels on each detector-half, respectively, is given by the integral over the pixel area. We thus introduce pixels of size $\Delta L$ and with effective loss-rate $\gamma = 1-(\text{fill-factor})\times(\text{quantum efficiency})$ and integrate  $P_C(x_1,x_2,\delta)$ over the interval $L_i = \left[-L/2+(i+\gamma/2)\Delta L,-L/2+(i+1-\gamma/2)\Delta L\right]$ for $x_1$ and $L_j$ over the same range for $x_2$, where $L$ is the total width of the (half)array; $P_C^{i j}(\delta) = \int_{L_i} dx_1 \int_{L_j} dx_2 \; P_C(x_1,x_2,\delta)$ is the coincidence probability between pixel $i$ and pixel $j$. The total coincidence probability is finally given by $P_C(\delta) = \sum_{i j} P_C^{i j}(\delta)$. Computing the visibility,
\begin{align}
V = \left|\frac{P_C(\delta \rightarrow \infty)-P_C(\delta = 0)}{P_C(\delta \rightarrow \infty)+P_C(\delta = 0)}\right|,
\end{align}
yields the plot [dashed curve shown in Fig. \ref{Figure3}(d)], which captures the overall trend with $R = \Delta L/(2\sqrt{2\ln 2}\sigma_\text{corr})$ and where we used the experimentally relevant parameters: $\gamma = 1- 0.09\times 0.78 =0.93$, $k_p = 2\pi/347\text{ nm}$, $L_c = 0.5\text{ mm}$.
\\}
{\bf{Photon number resolution approach for improved signal to noise.}}  
{{We can provide an estimate of the noise in each image by analysing the count statistics across pixels that are uniformly illuminated. For example, Fig.~\ref{Figure5}f has a region of abut 200 pixels that is illuminated by the central and uniform region of the PDC emission. Across these pixels we evaluate the ratio of the standard deviation of the square root of the mean count value to be 1.26, i.e. the camera pixels report counts that are $\sim1.3$x the shot noise limit. \\
Noise in these measurements has several origins: losses in the system, dark count noise on the camera, pixel fill-factor and effective photon loss on the camera due to the 5-10\% quantum efficiency. These losses will then affect photon coincidence images quadratically, i.e. the noise in photon pair counting is given by the product of the noise of the two individual photons.\\}}
In the presence of photon loss (or detectors with limited efficiency), distinguishing single-photon clicks from bunching and coincidence events can increase the precision of HOM-based sensing~\cite{Scott2020}.
The HOM signal is not constrained to the anti-diagonal ($k_x=-k_x$) of the JPD which indicates coincidence between a pixel and its coincidence partner; additional coincidence and bunching information is found through the JPD terms correlating a pixel with  neighbours of itself and of its coincidence partner  (the non-zero values of the sum and minus coordinates shown in Fig.~\ref{Figure3}) that arise due to a correlation point-spread function which, although matched to the pixel size as described in the main text, will spread across adjacent pixels due to its Gaussian-like distribution.
These latter terms allow us to harness the number-resolving advantage in the fundamental HOM experiment~\cite{Scott2020} as well as addressing array-specific noise contributions~\cite{Defienne2021}.
In order to account for different quality dips across the images we rescale the images based on image regions of constant coincidence counts. For example, we select with a mask, the inner or the outer regions of the cross and the inner or outer regions of the `UofG' lettering
These rescaled images are then combined according to the estimator for a common signal in multiple independent noisy channels---$\hat{\theta} = [\,\sum_j\sigma_j\,]  \sum_j (x_j/\sigma_j)$---to obtain a minimum-variance estimate~\cite{kay_fundamentals_1998}.
Here, $j$ denotes the anti-bunching and bunching images that are therefore summed together with relative weights given by the $\sigma_j$ values, i.e. the associated standard deviation in the counts that are used as a noise estimate and are computed in an image region of constant coincidence counts.\\


%

\end{document}